\documentclass[11pt]{article}

\usepackage[utf8]{inputenc}
\usepackage[T1]{fontenc}
\usepackage{lmodern}
\usepackage{geometry}
\usepackage{float}

\usepackage[justification=centering]{caption}

\usepackage{tabularx}
\usepackage{graphicx}
\usepackage{hyperref}
\usepackage{booktabs}
\usepackage{amsmath}
\usepackage{amssymb}
\usepackage{tabularx}
\usepackage{setspace}
\usepackage{svg}
\usepackage{array}
\usepackage{multirow}
\usepackage{ragged2e}
\newcolumntype{L}[1]{>{\RaggedRight\arraybackslash}p{#1}}
\newcolumntype{C}[1]{>{\Centering\arraybackslash}p{#1}}
\newcolumntype{R}[1]{>{\RaggedLeft\arraybackslash}p{#1}}

\title{Seasonality in the U.S. Housing Market: Post-Pandemic Shifts and Regional Dynamics}

\author{
  \href{https://orcid.org/0009-0009-5857-5783}{\includegraphics[scale=0.06]{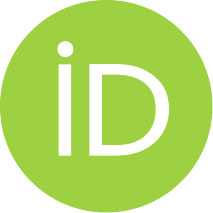}\hspace{1mm}Yihan Hu}$^{1}$\thanks{Correspondence: \texttt{Yh623@cam.ac.uk}; Tel.: +44~7526~543793}\\[0.8ex]
  \href{https://orcid.org/0009-0007-8964-7941}{\includegraphics[scale=0.06]{orcid.pdf}\hspace{1mm}Yifei Huang}$^{2}$,
  \href{https://orcid.org/0009-0001-9131-4632}{\includegraphics[scale=0.06]{orcid.pdf}\hspace{1mm}Weizhao Wang}$^{3}$\\[1ex]
  \small $^{1}$University of Cambridge, Cambridge, CB2 1TN, United Kingdom; \texttt{Yh623@cam.ac.uk}\\
  \small $^{2}$Northwestern University, Evanston, IL 60208, United States; \texttt{Yifeihuang2027@u.northwestern.edu}\\
  \small $^{3}$Johns Hopkins University, Baltimore, MD 21218, United States; \texttt{Wwang206@jh.edu}
}

\date{\today}

\begin{document}

\maketitle

\begin{abstract}
Seasonality has traditionally shaped the U.S. housing market, with activity peaking in spring--summer and declining in autumn--winter. However, recent disruptions, particularly post-COVID-19, raise questions about shifts in these patterns. This study analyzes housing market data (1991--2024) to examine evolving seasonality and regional heterogeneity. Using Housing Price Index (HPI), inventory and sales data from the Federal Housing Finance Agency and the U.S. Census Bureau, seasonal components are extracted via the X-13-ARIMA procedure, and statistical tests assess variations across regions. The results confirm seasonal fluctuations in prices and volumes, with recent shifts toward earlier annual peaks (March--April) and amplified seasonal effects. Regional variations align with differences in climate and market structure, while prices and sales volumes exhibit in-phase movement, suggesting thick-market momentum behaviour. These findings highlight key implications for policymakers, realtors and investors navigating post-pandemic market dynamics, offering insights into the timing and interpretation of housing market activities.
\end{abstract}


\section{Introduction}

Seasonality is a well-documented feature of housing markets. Particularly in the United States, home prices and sales volumes typically increase during the spring and summer months and decline in the fall and winter  \cite{Gourley2021}. These cyclical patterns, often referred to as ``hot'' and ``cold'' seasons, have traditionally been attributed to factors such as weather conditions, the school calendar, and the timing preferences of buyers and sellers. Understanding these patterns is not merely of academic interest; it holds practical value for optimizing household real estate decisions and guiding industry practice \cite{Seagraves2023}. For instance, sellers often achieve higher prices and faster transactions during the spring surge in demand, while buyers may benefit from greater bargaining power during the winter slowdown.

However, recent disruptions, particularly those stemming from the COVID-19 pandemic, have introduced new uncertainty to housing seasonality \cite{Malpezzi2023}. The pandemic and its associated economic shocks disrupted traditional market behavior, leading to unseasonal surges or delays in housing activity during 2020 \cite{Wang2024}. By 2021--2022, analysts observed that some established seasonal patterns had either weakened or shifted, with housing demand reaching historic highs and lows in atypical months. These changes raise a critical question: have the long-standing seasonal rhythms of the U.S. housing market fundamentally evolved in the post-pandemic period?

Additionally, regional disparities in housing cycles have become more pronounced. Variations in climate, regional economic conditions, and migration trends mean that one region’s ``spring boom'' may occur at a different time of year than another’s \cite{Kuepers2023}. Despite these observations, systematic research on how seasonality has shifted across U.S. regions in the wake of the pandemic remains limited. Existing studies largely focus on pre-2020 patterns or narrow aspects, such as seasonal adjustment methods or single-city dynamics, leaving gaps in understanding the broader regional and temporal impacts of recent disruptions \cite{Martin2021,Kumar2025,BurtonJones2021}.

This study seeks to address these gaps by providing a comprehensive analysis of U.S. housing market seasonality with a particular focus on the post-pandemic period and regional heterogeneity. Specifically, we investigate whether the timing and intensity of seasonal peaks and troughs have shifted following the pandemic and quantify the extent to which the magnitude of seasonality varies across regions. These findings are crucial, as uneven shifts in seasonality could have significant implications for housing affordability, inventory management, and policy effectiveness at the regional level.

In summary, this study makes two key contributions. First, it identifies a forward shift in the timing of seasonal housing market peaks after the COVID-19 shock, reflecting an earlier onset of the ``hot season'' in many regions. Second, it provides a detailed comparison of seasonal patterns across U.S. regions over recent decades, offering insights into the geographical heterogeneity of these changes. Methodologically, this study leverages a large dataset of monthly housing indicators (prices, inventory, and sales) spanning 1991 to 2024. Seasonal components are extracted using the X-13ARIMA-SEATS decomposition, and formal statistical tests and panel regression analyses (including region-by-season interactions) are conducted to validate the presence and variation of seasonality. These findings enhance our understanding of evolving market dynamics and provide actionable insights for policymakers, real estate professionals, and market participants navigating the post-pandemic U.S. housing market.

\section{Literature review}

\subsection{Seasonal patterns and the thick-market effect}

Early studies demonstrate that housing markets exhibit predictable seasonal cycles \cite{Nong2024,RoedLarsen2024,Fracz2023}. Prices and transaction volumes tend to peak in spring and summer, followed by a decline in fall and winter. These patterns have been attributed to factors such as weather, school calendars, and buyer--seller preferences \cite{Etukudor2020}. A critical insight from these studies is the ``thick-market effect,'' which explains how heightened activity during peak seasons reinforces itself \cite{GhaniReed2022}. Increased buyer density improves sellers’ bargaining power and accelerates inventory turnover, creating a self-reinforcing loop of rising prices and transaction volumes during peak periods.

Empirical evidence supports this dynamic, revealing that housing markets deviate from classical supply--demand mechanisms during peak seasons. For instance, Novy-Marx \cite{NovyMarx2009} showed that seasonal price--volume increases reflect a shift in the entire supply--demand equilibrium, rather than a simple price adjustment. While these findings provide a strong theoretical foundation, much of this research focuses on pre-2020 patterns and assumes stable seasonal dynamics over time. Few studies have investigated whether recent disruptions, such as the pandemic, have altered these established mechanisms or whether seasonal peaks have shifted temporally \cite{Liu2021}. This study builds on these theoretical frameworks by exploring how seasonality has evolved in both timing and magnitude in the post-pandemic era.

\subsection{Post-pandemic changes in seasonality}

The COVID-19 pandemic introduced unprecedented disruptions to housing market seasonality, challenging traditional cyclical patterns. McNamara \cite{McNamara2024} finds that since 2020, the U.S. housing market has exhibited amplified seasonal amplitudes, with stronger price growth during high-demand months and steeper declines during low-demand periods. Such deviations have exposed limitations in standard seasonal adjustment models (e.g., X-12/X-13), which may under-correct for the intensified fluctuations observed post-2020. For example, the summer surges and winter lulls of 2020--2021 were far more pronounced than in previous years, likely reflecting pandemic-induced shifts in consumer behavior, economic uncertainty, and policy interventions.

Moreover, recent studies suggest that the timing of seasonal peaks has shifted \cite{Kishore2020}. Traditional market activity typically reached its zenith in May or June; however, analyses of post-2020 data indicate that peak activity now occurs earlier, often in March or April. This shift may be attributed to changes in buyer urgency, greater adoption of technology facilitating year-round market participation, or lingering effects of the pandemic that reshaped housing demand patterns \cite{White2021}. While these studies highlight significant changes, they often focus on national trends and lack a comprehensive analysis of how these shifts vary across regions. To address this gap, this research examines both the timing and amplitude of seasonal changes across different U.S. regions using updated data from 1991 to 2024.

\subsection{Regional heterogeneity and drivers of seasonality}

Seasonality in housing markets is not uniform across regions, with significant geographic variation influenced by local climate, economic conditions, and demographics. Budikova et al.\ \cite{Budikova2022} documented that colder regions, such as the Northeast and Midwest, exhibit more pronounced seasonal swings, while warmer areas, such as the South Atlantic and Pacific Coast, experience milder fluctuations. These findings underscore the importance of region-specific seasonal adjustments to accurately capture local housing cycles.

In addition to geographic factors, transaction composition and behavioral dynamics play a critical role in shaping seasonal patterns. Hattapoglu \cite{Hattapoglu2021} observed that during peak seasons, a higher proportion of transactions involve larger homes or those in desirable neighborhoods, driving average prices upward. Behavioral preferences also reinforce these trends; for instance, Røed Larsen \cite{RoedLarsen2024} identified a persistent ``December discount'' in Norwegian housing markets, where thin winter markets lead to lower prices. Similar dynamics are evident in the U.S., where peak-season activity amplifies both prices and volumes, reflecting a self-reinforcing demand cycle.

Despite these insights, the literature remains fragmented in its treatment of regional variation, with most studies focusing on single-city or national-level analyses. Furthermore, little attention has been given to how local seasonal patterns have evolved in the wake of the pandemic. This study addresses these gaps by systematically analyzing regional differences in seasonal intensity and timing, providing new evidence on how one region’s ``off-season'' can coincide with another’s peak. By quantifying these interactions, this research contributes to a more nuanced understanding of housing market seasonality across geographic and temporal dimensions.

\section{Methodology}

This study employs a quantitative approach to examine seasonal patterns in the U.S. housing market, focusing on post-pandemic shifts and regional heterogeneity. The analysis is based on monthly housing market data spanning 1991--2024, with seasonal components isolated using statistical decomposition methods. This section outlines the research design, data sources, and analytical techniques used to address the research questions.

\subsection{Data sources and variables}

The study utilizes publicly available, high-frequency housing market data from the U.S. Census Bureau and the Federal Housing Finance Agency (FHFA). The dataset spans January 1991 to December 2024, covering critical economic events such as the 2008 financial crisis and the COVID-19 pandemic. Table~\ref{tab:data_sources} shows the data descriptions and sources.

\begin{table}[H]
    \centering
    \caption{Data sources and variables}
    \label{tab:data_sources}
    \begin{tabular}{L{2.7cm}L{6.5cm}L{3.5cm}L{3.6cm}}
        \toprule
        \textbf{Variable} & \textbf{Description} & \textbf{Source} & \textbf{Purpose} \\
        \midrule
        Housing Price Index (HPI) & Weighted repeat-sales index measuring changes in single-family home prices. & Federal Housing Finance Agency (FHFA) & Captures overall price trends and seasonal fluctuations in the housing market. \\
        Houses Sold & Monthly data on the number of new houses sold. & U.S. Census Bureau & Reflects demand intensity and transaction volumes. \\
        Houses for Sale & Monthly inventory of available houses for sale. & U.S. Census Bureau & Evaluates market supply and its seasonal variations. \\
        Median Sale Price & Median price of houses sold in a given month. & U.S. Census Bureau & Provides insights into typical buyer and seller behavior while minimizing sensitivity to outliers. \\
        Average Sale Price & Average price of houses sold in a given month. & U.S. Census Bureau & Complements the median price by providing additional context on price distribution. \\
        Months’ Supply & Ratio of inventory to sales, indicating how long the current inventory would last at the current sales rate. & Derived from ``Houses Sold'' and ``Houses for Sale'' & Measures market balance and highlights the ``hotness'' of the market during different seasons. \\
        Construction Stage Data & Categorization of houses as ``under construction,'' ``completed,'' or ``not started.'' & U.S. Census Bureau & Provides insights into supply-side dynamics and construction timing relative to seasonal demand. \\
        Regional Price Data & Median and average sale prices by census division. & U.S. Census Bureau & Enables comparison of seasonal trends across different U.S. regions. \\
        Type of Financing & Includes FHA loans, VA loans, and conventional financing. & U.S. Census Bureau & Explores buyer behavior and financing trends in response to market changes. \\
        FHFA House Price Index & Comprehensive house price index series measuring single-family home values. & Federal Housing Finance Agency (FHFA) & Provides a long-term view of housing price fluctuations and seasonality. \\
        \bottomrule
    \end{tabular}
\end{table}

\subsection{Analytical framework}

\subsubsection{Seasonal decomposition using X-13ARIMA-SEATS}

\begin{figure}[ht]
    \centering
    \includegraphics[width=0.8\textwidth]{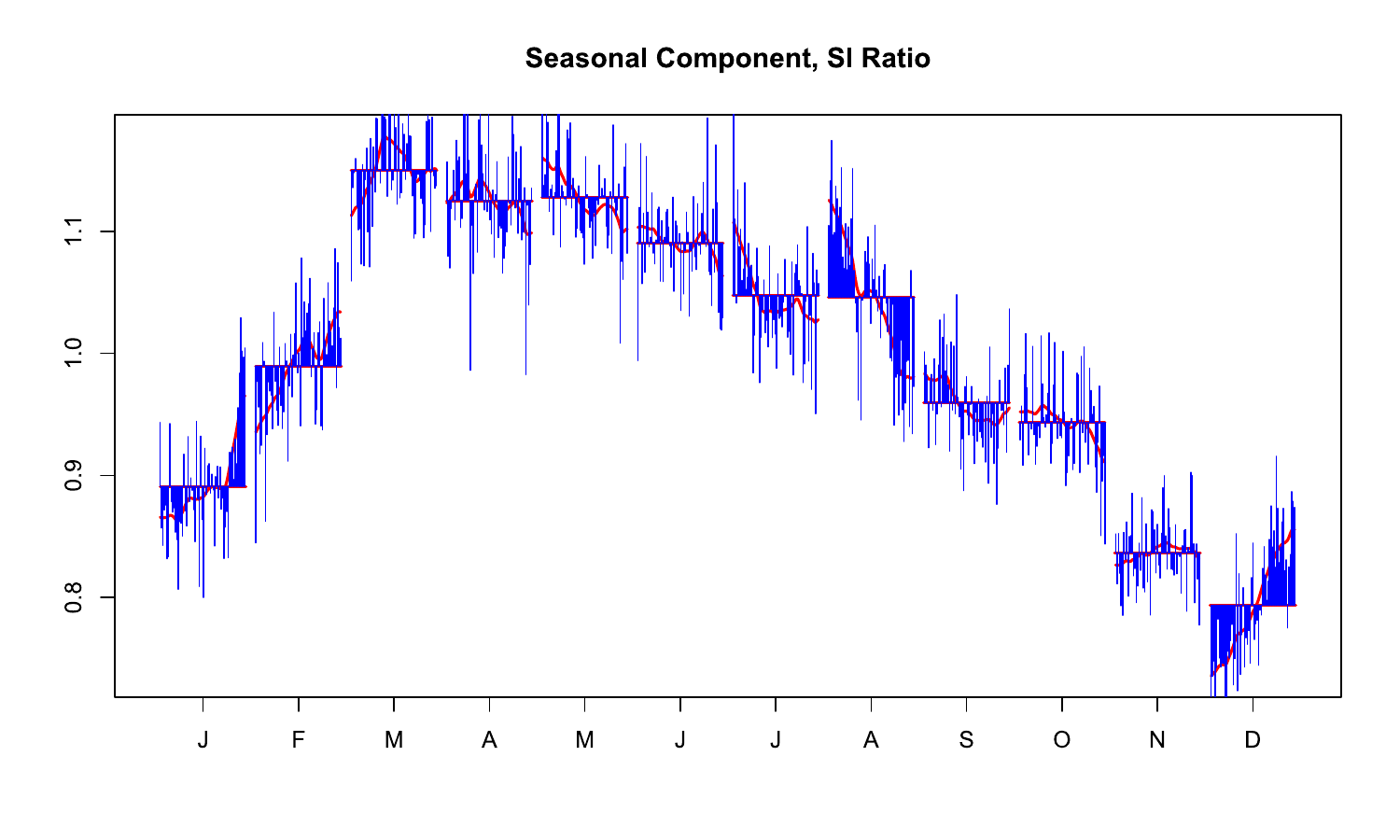}
    \caption{Illustration of X-13ARIMA-SEATS seasonal decomposition.}
    \label{fig:x13}
\end{figure}

To isolate and analyze the seasonal components, the study employs the X-13ARIMA-SEATS methodology (Figure~\ref{fig:x13}), a widely used time-series decomposition technique developed by the U.S. Census Bureau. This method decomposes each time series into three components:
\begin{equation}
    y_t = T_t + S_t + I_t,
\end{equation}
where $T_t$ represents the trend--cycle component capturing long-term movements, $S_t$ is the seasonal component capturing predictable intra-year fluctuations, and $I_t$ represents the irregular component accounting for random noise.

By separating these components, the method ensures a structured understanding of the cyclical patterns in the housing market. The X-13ARIMA-SEATS process involves several key steps, including pre-adjustment to remove outliers and calendar effects, fitting a Seasonal Auto-Regressive Integrated Moving Average (SARIMA) model, and performing diagnostic checks such as the Ljung--Box Q-test to confirm the randomness of residuals \cite{Corona2024}. This decomposition enables the study to distinguish real seasonality from random noise effectively.

\subsubsection{Statistical tests for seasonality}

To formally validate the presence and significance of seasonality, this study employs multiple statistical tests. The parametric F-test is used to assess whether monthly variations in housing prices and sales turnover are statistically significant. This test evaluates the null hypothesis that all monthly means are equal, with a significant F-statistic indicating strong seasonality. 

To account for potential non-normality in the data, the nonparametric Kruskal--Wallis test is also employed. This test ranks observations across months and evaluates whether distributions differ significantly without assuming normality. Additionally, a two-way ANOVA is conducted to examine interactions between months (seasonal effects) and years (long-term trends), providing insights into whether seasonal patterns evolve over time. The study also applies the Chow test to detect structural breaks in seasonality, identifying whether key economic events, such as the 2008 financial crisis or the COVID-19 pandemic, have caused significant shifts in the seasonal structure of the housing market.

\subsubsection{Regression analysis for regional variations}

To quantify regional differences in seasonal patterns, a linear regression model with interaction terms is employed:
\begin{equation}
    \text{HPI}_{it} = \alpha + \beta S_{it} + \gamma R_i + \delta (S_{it} \times R_i) + \varepsilon_{it},
\end{equation}
where $\text{HPI}_{it}$ denotes the non-seasonally adjusted HPI for region $i$ in month $t$, $S_{it}$ represents the seasonal component extracted by X-13ARIMA-SEATS, $R_i$ is a categorical variable for the nine census divisions, and $S_{it} \times R_i$ is the interaction term capturing region-specific seasonality. The interaction term provides insights into how seasonal trends differ across regions, addressing a critical gap in the literature on geographic heterogeneity in housing market dynamics.

Finally, robustness checks are conducted to ensure the reliability of the results. Sensitivity analyses are performed to test the impact of excluding outliers and extreme observations. Alternative regression models, including fixed-effects and random-effects specifications, are compared to verify the consistency of results. Residual diagnostics, such as autocorrelation checks using the Ljung--Box Q-test and autocorrelation function (ACF) plots, are employed to confirm the adequacy of the model.

\section{Results}

\subsection{Evidence of seasonality}

The analysis reveals clear and significant seasonal patterns in the U.S. housing market. Using the X-13ARIMA-SEATS methodology, distinct cyclical components are identified in key market indicators, including the HPI, the number of houses sold, and the inventory of houses for sale.

\begin{figure}[ht]
    \centering
    \includegraphics[width=0.8\textwidth]{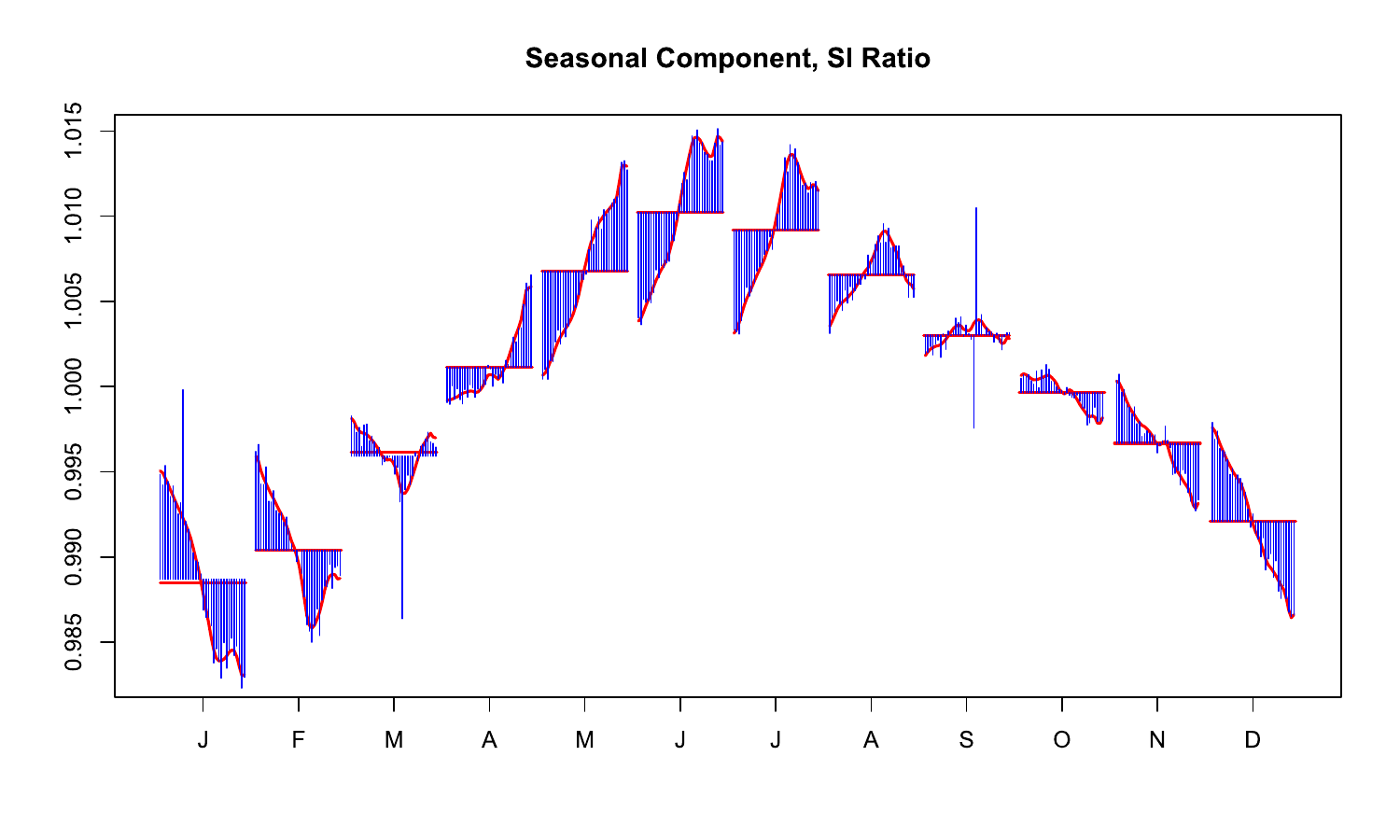}
    \caption{Seasonal component of the Housing Price Index.}
    \label{fig:hpi_seasonal}
\end{figure}

Figure~\ref{fig:hpi_seasonal} illustrates the seasonal component of the HPI. Prices exhibit predictable fluctuations, with peaks consistently occurring during the summer months (June--August) and troughs during the winter months (December--February). The seasonal index for the HPI remains above 1.005 from April to August, reflecting increased market activity during these months, while it falls below 0.995 in December and January, indicating a slowdown in market activity.

Similarly, the seasonal components of ``Houses Sold'' and ``Houses for Sale'' show analogous patterns. Transaction volumes rise sharply from early spring and peak in the summer before declining steadily in the fall and reaching their lowest levels in winter. Inventory levels follow a similar seasonal trajectory, with higher availability in spring and summer and reduced supply during winter months. These findings underline the existence of strong seasonality across multiple dimensions of the housing market.

\subsection{Statistical validation of seasonality}

To formally validate the observed seasonality, several statistical tests are conducted. The F-test results demonstrate significant monthly variations across all key variables. For instance, the F-statistic for the HPI is 61.63 ($p < 0.001$), indicating substantial differences in monthly means. This confirms that housing prices are not uniformly distributed throughout the year, but instead follow a systematic seasonal pattern.

The Kruskal--Wallis test, a nonparametric alternative to the F-test, further corroborates these findings. With a test statistic of 311.91 ($p < 0.001$), the Kruskal--Wallis test confirms that the monthly distributions of housing prices and transaction volumes differ significantly. Importantly, this test does not assume normality in the data, making it particularly robust for variables like ``Houses Sold,'' which may be affected by outliers or non-normal distributions.

The results of a two-way ANOVA provide additional insights. Both the main effects of Month ($p < 0.001$) and Year ($p < 0.001$) are statistically significant, indicating that seasonal patterns persist across years but vary in magnitude. Furthermore, the significant Month $\times$ Year interaction term ($p < 0.001$) suggests that the intensity and timing of seasonal fluctuations have evolved over time.

Table~\ref{tab:stats} summarises the key test statistics.

\begin{table}[H]
    \centering
    \caption{Statistical results for seasonality tests}
    \label{tab:stats}
    \begin{tabular}{l l c c c c c}
        \toprule
        \textbf{Test type} 
        & \textbf{Test / Term} 
        & \textbf{df} 
        & \textbf{Sumsq} 
        & \textbf{Meansq} 
        & \textbf{Statistic} 
        & \textbf{$p$-value} \\
        \midrule
        \multirow{2}{*}{Seasonality tests} 
        & Parametric F-test     & --  & --        & --        & $F = 61.63$  & $< 0.001^{***}$ \\
        & Kruskal--Wallis       & --  & --        & --        & $H = 311.91$ & $< 0.001^{***}$ \\
        \midrule
        \multirow{3}{*}{Two-way ANOVA (HPI)} 
        & Month                 & 11  & 4788.18    & 435.29    & 22.41        & $< 0.001^{***}$ \\
        & Year                  & 33  & 2837411.60 & 85982.17  & 4426.58      & $< 0.001^{***}$ \\
        & Residuals             & 360 & 6992.66    & 19.42     & --           & --              \\
        \bottomrule
    \end{tabular}
\end{table}

\subsection{Evolution of seasonal patterns}

\begin{figure}[H]
  \centering
  \includegraphics[width=0.48\textwidth]{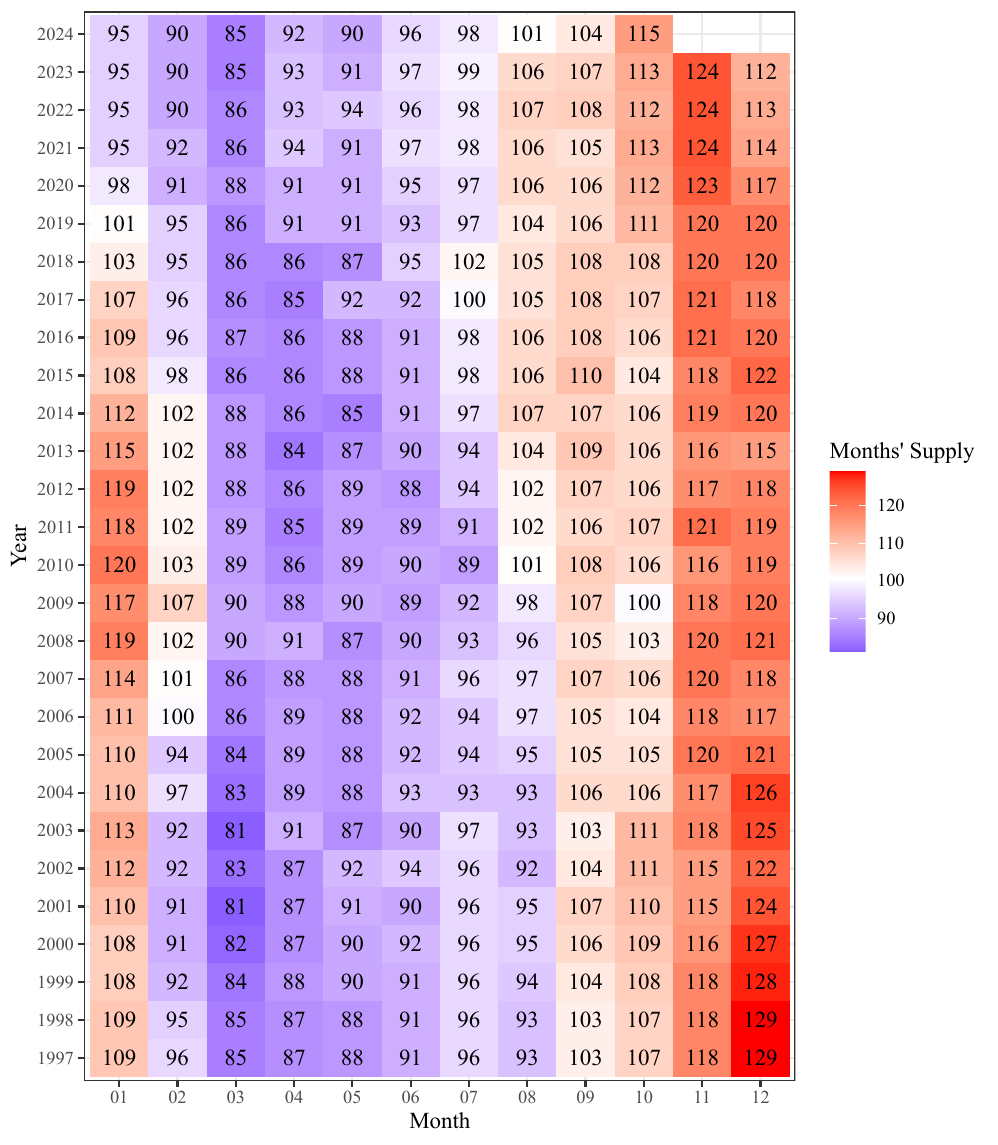}
  \hfill
  \includegraphics[width=0.48\textwidth]{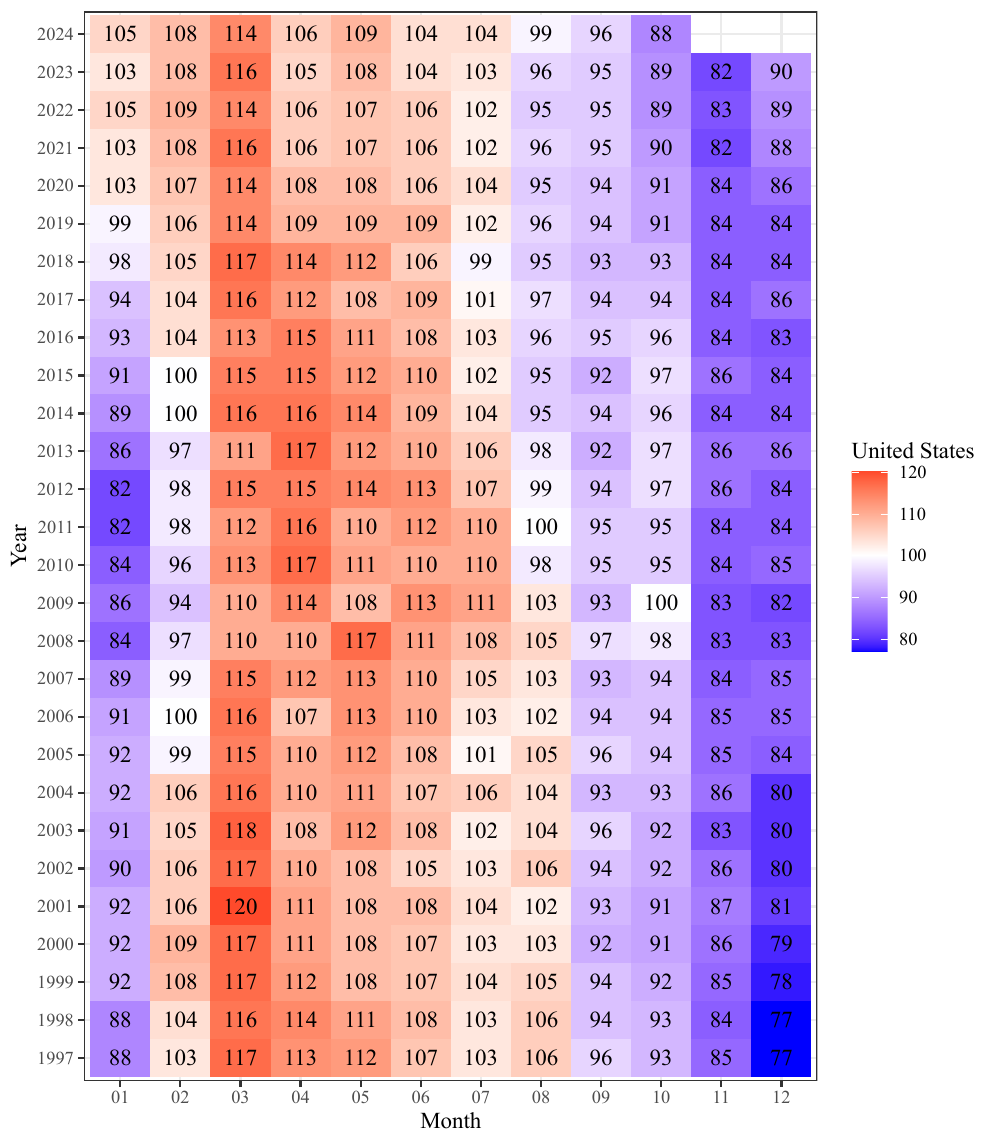}
  \caption{Changes in the timing of seasonal peaks in housing market indicators.
  \newline Left: Months' Supply; Right: United States index.}
  \label{fig:seasonal_shift}
\end{figure}

The analysis indicates a notable shift in the timing and intensity of seasonal patterns in recent years. Figure~\ref{fig:seasonal_shift} highlights changes in the seasonal peaks and troughs for housing sales activity. Historically, the peak in housing sales occurred in May or June. However, in the post-2020 period, this peak has shifted to earlier months, such as March and April. Similarly, inventory recovery, which traditionally began in March, now starts as early as January or February.

The variable ``Months’ Supply'' provides further evidence of these shifting patterns. Figure~\ref{fig:seasonal_shift} shows that the peak inventory levels, which historically occurred in December, have moved to October or November in recent years. These findings suggest that the seasonal cycle of the U.S. housing market is gradually shifting forward, potentially reflecting changes in consumer behavior, macroeconomic conditions, or market dynamics following the COVID-19 pandemic.

\subsection{Regional variations in seasonality}

\begin{figure}[H]
  \centering
  \includegraphics[width=0.8\textwidth]{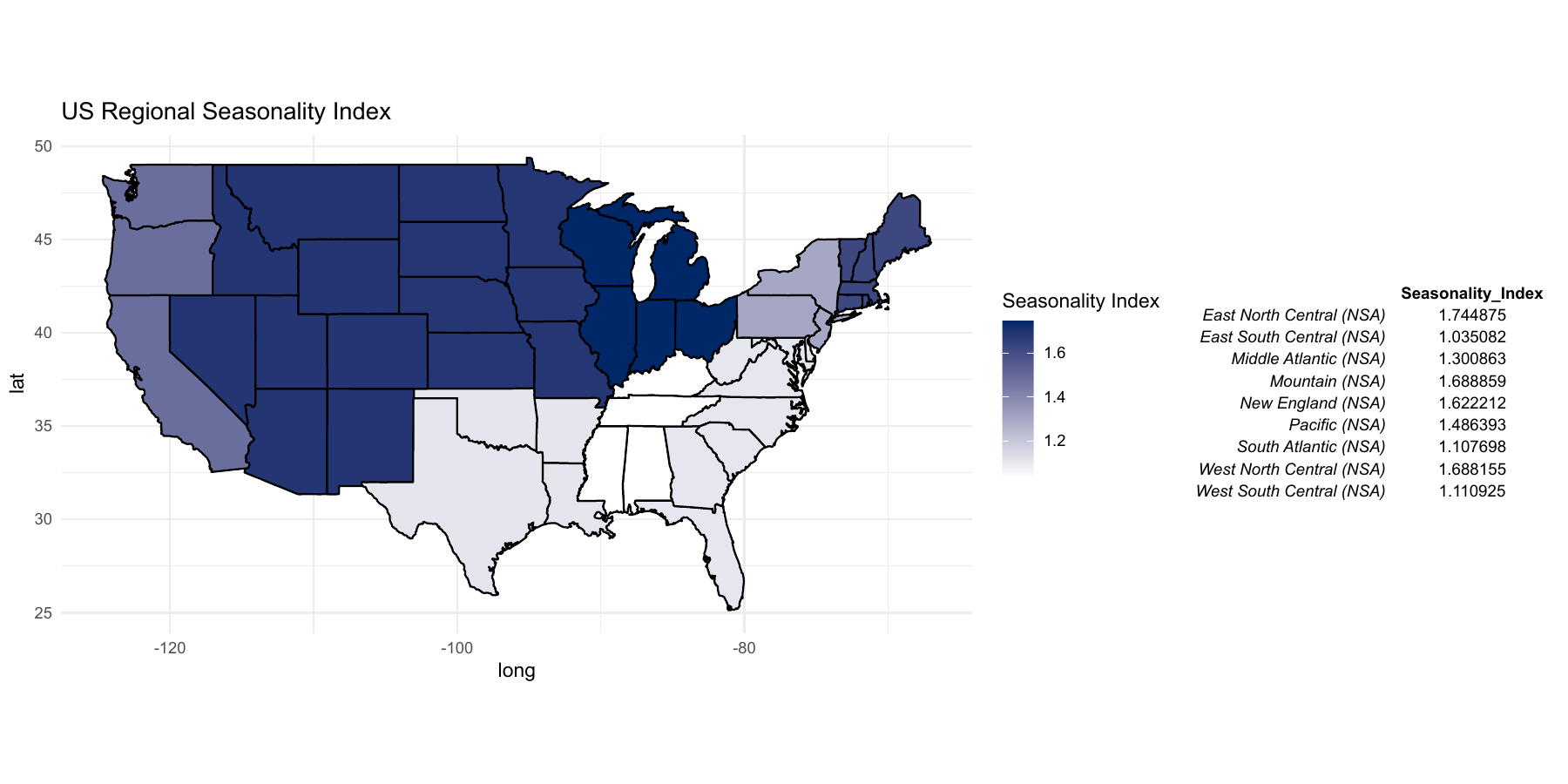}
  \caption{Regional variation in the intensity of housing market seasonality.}
  \label{fig:regional}
\end{figure}

Seasonality in the U.S. housing market varies significantly across regions, as shown in Figure~\ref{fig:regional}. The Northeast and Midwest exhibit the most pronounced seasonal fluctuations, with sharper declines in housing activity during the winter months. These regions experience harsher weather conditions, which likely discourage transactions and limit construction activity during colder months.

In contrast, the South and West regions demonstrate relatively weaker seasonality. The milder climates in these areas allow for year-round construction and market activity, reducing the amplitude of seasonal fluctuations. For example, in the South Atlantic region, housing prices and sales volumes show a smaller seasonal drop during winter compared to the Northeast or Midwest.

Regression analysis confirms these regional differences. The interaction term between seasonality and region is significant ($p < 0.001$), indicating that the impact of seasonality on housing prices varies geographically. For instance, the South Atlantic region shows a positive interaction effect, suggesting that seasonal dynamics amplify price increases more strongly in this region compared to the national average. Conversely, the West North Central region exhibits a negative interaction effect, indicating weaker seasonal impacts on housing prices.

\begin{table}[H]
    \centering
    \caption{Regression results for seasonality--region interactions}
    \label{tab:reg}
    \begin{tabular}{L{5cm} r r r l}
        \toprule
        \textbf{Term} & \textbf{Estimate} & \textbf{Std. Error} & \textbf{t-value} & \textbf{$p$-value} \\
        \midrule
        Seasonality $\times$ Region East North Central & -8.72 & 2.45 & -3.56 & $0.000^{***}$ \\
        Seasonality $\times$ Region East South Central &  5.42 & 2.99 &  1.82 & 0.007 \\
        Seasonality $\times$ Region Middle Atlantic     & -2.41 & 2.71 & -0.89 & 0.373 \\
        Seasonality $\times$ Region Mountain            & -0.34 & 2.25 & -0.15 & 0.880 \\
        Seasonality $\times$ Region New England         & -3.41 & 2.37 & -1.44 & 0.150 \\
        Seasonality $\times$ Region Pacific             & -3.17 & 2.52 & -1.25 & 0.210 \\
        Seasonality $\times$ Region South Atlantic      &  7.76 & 2.65 &  2.93 & $0.003^{**}$ \\
        Seasonality $\times$ Region West North Central  & -5.26 & 2.40 & -2.19 & $0.029^{*}$ \\
        Seasonality $\times$ Region West South Central  &  0.60 & 2.77 &  0.22 & 0.828 \\
        \bottomrule
    \end{tabular}

    \vspace{0.2cm}
    {\small Note. Data source: HPI data, 1991--2004.\\
    $^{*}p < 0.05$, $^{**}p < 0.01$, $^{***}p < 0.001$.}
\end{table}

\subsection{Robustness of results}

The robustness of the results is evaluated using a series of sensitivity analyses and diagnostic tests to ensure the findings are reliable and not influenced by data anomalies, model misspecification, or structural changes in the housing market.

To determine whether outliers in housing price data influence the regression results, observations in the top and bottom 1\% of the HPI distribution are excluded, and the models are re-estimated using the truncated dataset. This analysis reveals that the exclusion of extreme observations has no substantive impact on the findings. Key coefficients, such as the seasonal effect for July and January, change only marginally. For example, the coefficient for July decreases slightly from 0.062 ($p < 0.001$) to 0.061 ($p < 0.001$), while the coefficient for January changes from -0.058 ($p < 0.001$) to -0.057 ($p < 0.001$). Similarly, regional effects, such as the premium for the Pacific region, remain stable at approximately 0.198. These results confirm that the findings are not driven by extreme observations and reflect broader trends in the housing market.

In addition, residual diagnostics are conducted to evaluate whether the regression model meets the assumptions of linear regression, including homoscedasticity, normality, and the absence of autocorrelation. The Breusch--Pagan test for heteroscedasticity yields a $p$-value of 0.21, indicating no evidence of heteroscedasticity. The Shapiro--Wilk test for normality returns a $p$-value of 0.13, suggesting that the residuals approximate a normal distribution. Furthermore, the Ljung--Box Q-test for autocorrelation shows no significant results ($p > 0.10$), confirming that residuals are uncorrelated over time. Visual diagnostic tools, including autocorrelation function (ACF) plots, histograms of residuals, and Q--Q plots, visually support these findings by showing no significant deviations from normality or randomness. Collectively, these results affirm that the regression model satisfies the assumptions necessary for valid statistical inference.

\section{Discussion}

This study provides a comprehensive analysis of seasonal dynamics and regional heterogeneity in the U.S. housing market, with a particular focus on post-pandemic shifts. By leveraging data from 1991 to 2024 and employing advanced statistical techniques such as X-13ARIMA-SEATS decomposition, the findings reveal significant changes in the timing and intensity of seasonality as well as pronounced regional variations.

The results reveal that while seasonality remains a prominent feature of the U.S. housing market, its patterns have undergone significant shifts in the post-pandemic period. The findings indicate that the timing of seasonal peaks has advanced, with housing market activity now reaching its zenith as early as March or April, compared to the traditional peak in May or June. This shift aligns with observations by Duca et al.\ \cite{Duca2021}, who documented accelerated seasonal cycles and amplified fluctuations in housing prices and sales volumes since the pandemic. These changes are likely driven by a combination of altered consumer behavior, technological adoption facilitating year-round transactions, and lingering macroeconomic adjustments. This evidence challenges the assumption of stable seasonality in traditional housing market models and highlights the need for adaptive frameworks.

Regional heterogeneity emerges as a critical dimension of housing market seasonality. Colder regions, such as the Northeast and Midwest, exhibit more pronounced seasonal variations, with sharper declines in market activity during winter months. This finding corroborates the work of Mahmud et al.\ \cite{Mahmud2021}, who attributed these pronounced fluctuations to adverse weather conditions that limit transactions and construction activity. In contrast, warmer regions, such as the South Atlantic and Pacific Coast, display milder seasonal swings, reflecting year-round market participation. These results extend prior research by providing a detailed comparative analysis of seasonal dynamics across regions, emphasizing how geographic and climatic factors shape local housing cycles.

The forward shift in seasonal peaks observed in this study also reflects structural changes in market behavior. Historically, the ``spring boom'' in housing activity was closely tied to school calendars and weather-driven preferences. However, the earlier onset of peak activity suggests that these traditional drivers may now be complemented by new factors, such as heightened buyer urgency and changing migration patterns. This shift aligns with Moser et al.’s \cite{Moser2022} findings on increased demand for larger homes in suburban areas during the pandemic, as remote work enabled greater geographic flexibility. By shifting the timing of demand, these behavioral changes may have permanently altered the seasonal rhythm of the housing market.

Additionally, the study confirms the self-reinforcing nature of the thick-market effect, as described by Hattapoglu and Hoxha \cite{Hattapoglu2021}. During peak months, higher buyer density contributes to increased bargaining power for sellers, driving both prices and transaction volumes upward. However, the findings suggest that this effect is now more pronounced in certain regions, such as the South Atlantic, where seasonal price increases significantly exceed the national average. This regional amplification of the thick-market effect underscores the importance of understanding local market dynamics, as a national-level analysis may obscure critical geographic differences.

The study also highlights the limitations of traditional seasonal adjustment models, such as X-12 and X-13ARIMA-SEATS, when applied to post-pandemic housing data. These models assume relatively stable seasonal amplitudes and timing, which may not accurately capture the amplified fluctuations and forward shifts observed in recent years. For example, Mahmud et al.\ \cite{Mahmud2021} noted that these models often under-correct for intensified seasonal peaks, leading to overestimation of off-season activity. The present study reinforces this critique by demonstrating that unadjusted seasonal components provide a clearer picture of housing market dynamics, particularly in regions with strong climatic effects.

In comparing these results to pre-pandemic studies, it is evident that the resilience of seasonality has persisted despite external disruptions. Earlier works, such as Novy-Marx \cite{NovyMarx2009}, emphasized the stability of seasonal cycles as a core feature of housing markets. However, the current findings suggest that while the underlying seasonal structure remains intact, its expression has evolved in both timing and intensity. This evolution highlights the adaptability of housing markets to external shocks, such as the COVID-19 pandemic, and underscores the importance of continuously monitoring these dynamics to inform both policy and practice.

\section{Conclusions}

This study demonstrates that seasonality remains a defining feature of the U.S. housing market, yet its nature is evolving. Using data from 1991 to 2024, the analysis confirms the persistence of the spring/summer boom and winter slowdown, while highlighting shifts in the timing and intensity of these patterns. Notably, seasonal peaks appear to have shifted earlier, with evidence suggesting that March and April are increasingly critical months in many markets. Regional heterogeneity further underscores the dynamic nature of seasonality, as colder regions exhibit more pronounced seasonal swings compared to milder climates.

These findings carry important implications for market participants and policymakers. First, timing strategies for buyers and sellers must adapt to the shifting seasonal window. Sellers aiming to capitalize on peak demand may need to list earlier in the year, while buyers seeking competitive opportunities may benefit from entering the market before traditional peaks. For example, in colder regions, listing in May rather than July can attract significantly more traffic. Second, policymakers and analysts should recalibrate seasonal adjustment models to account for these changes. Outdated adjustment methods may misinterpret trends, understating spring price growth or overstating summer slowdowns. Recognising regional diversity in seasonality can also improve policy targeting, such as adjusting the timing of initiatives to address housing supply shortages or managing loan demand surges during seasonal peaks.

Despite external disruptions such as the COVID-19 pandemic, the resilience of seasonality highlights its importance in housing market dynamics. However, stakeholders should be cautious about potential market instabilities. If widespread behaviour shifts occur in response to perceived seasonal changes, such as earlier buying or selling, it could amplify booms or slumps. Continuous monitoring of seasonal trends, along with research into micro-level behavioural changes, remains essential to understanding and mitigating these risks.

\section{Limitations and future work}

While this study provides valuable insights into the evolving nature of housing market seasonality, it is subject to several limitations. First, the analysis relies on regional aggregate data, which may mask finer local variations. Future research could focus on metropolitan-level or property-level data to uncover sharper idiosyncrasies, such as whether individual homes now achieve higher premiums in March compared to a decade ago.

Second, the methodological approach, based on X-13ARIMA-SEATS and classical statistical tests, assumes relatively smooth and stable seasonal patterns. This might overlook nonlinear or short-term fluctuations, particularly during periods of significant economic or policy shifts. Advanced time-varying parameter models, such as state-space approaches, could better capture gradual or abrupt changes in seasonality over time.

Third, the study primarily focuses on the for-sale housing market, without explicitly addressing seasonality in rental or commercial real estate markets. Given that sectors like apartment rentals exhibit distinct seasonal turnover patterns, future research could explore the interaction between rental and for-sale markets to provide a more comprehensive understanding of housing dynamics.

Lastly, while the study identifies shifts in seasonality and their potential drivers, it does not quantitatively model these factors. The role of school calendars, credit conditions, and demographic trends, among other factors, could be formally tested using structural models, such as equilibrium frameworks incorporating seasonally varying search intensity (e.g., Ngai \& Tenreyro, 2014). These extensions would enable a deeper exploration of the underlying mechanisms driving seasonal changes, enriching the understanding of housing market seasonality in a rapidly evolving environment.

\end{document}